\documentclass[iop,revtex4]{emulateapj}

\usepackage{natbib}
\usepackage{longtable}
\usepackage{rotating}
\usepackage{epstopdf}
\usepackage{graphicx}
\usepackage{subfigure}
\usepackage[usenames,dvipsnames]{color}
\usepackage{ifthen}

\def\cm2{cm$^{-2}$}

\def\nh3{NH$_3$}
\def\n2h{N$_2$H$^+$}

\def\13co{$^{13}$CO}
\def\c18o{C$^{18}$O}
\def\hc3n{HC$_3$N}
\def\h2{H$_2$}
\def\nh{n(H$_2$)}

\def\lc{\>\> ,}

\def\rsun{\ifmmode {\rm R}_{\mathord\odot}\else $R_{\mathord\odot}$\fi}
\def\msun{\ifmmode {\rm M}_{\mathord\odot}\else $M_{\mathord\odot}$\fi}
\def\lsun{\ifmmode {\rm L}_{\mathord\odot}\else $L_{\mathord\odot}$\fi}

\begin{document}

\slugcomment{}

\shorttitle{Constraining Molecular Cloud Thickness} \shortauthors{Qian, Li, Offner, Pan}

\title{A New Method for Constraining Molecular Cloud Thickness: A study of Taurus, Perseus and Ophiuchus}

\author{ Lei Qian \altaffilmark{1}\footnote{LQ analysed the observational data and wrote the paper except the hydrodynamical simulation part.}, Di Li \altaffilmark{1} \altaffilmark{2} \altaffilmark{4}\footnote{DL proposed the ideas to study the influence of thickness on the CVD. He also contributed to the paper text.}, Stella Offner \altaffilmark{3}\footnote{SO provided the hydrodynamical simulation, performed synthetic CO observations, wrote the hydrodynamical simulation part of the paper. She also made overall modification to the paper.} and Zhichen Pan \altaffilmark{1}\footnote{ZP helped to analyse the observational data.}}
\affil{} \altaffiltext{1} {National Astronomical Observatories,
Chinese Academy of Sciences, Beijing, 100012, China}
\altaffiltext{2} {Key Laboratory of Radio Astronomy, Chinese Academy of Sciences, Beijing, China}
\altaffiltext{3} {Department of Astronomy, University of Massachusetts, Amherst, MA, USA}
\altaffiltext{4} {Space Science Institute, Boulder, Colorado, USA}

\begin{abstract}
The core velocity dispersion (CVD) is a potentially useful tool for studying the turbulent velocity field of molecular clouds. CVD is based on centroid velocities of dense gas clumps, thus is less prone to density fluctuation and reflects more directly the cloud velocity field. Prior work demonstrated that the Taurus molecular cloud CVD resembles the well-known Larson's linewidth-size relation of molecular clouds. In this work, we studied the dependence of the CVD on the line-of-sight thickness of molecular clouds, a quantity which cannot be measured by direct means. We produced a simple statistical model of cores within clouds and analyzed the CVD of a variety of hydrodynamical simulations. We show that the relation between the CVD and the 2D projected separation of cores ($L_{2D}$) is sensitive to the cloud thickness. When the cloud is thin, the index of  CVD-$L_{2D}$ relation ($\gamma$ in the relation CVD$\sim L_{2D}^{\gamma}$) reflects the underlying energy spectrum ($E(k)\sim k^{-\beta}$) in that $\gamma\sim(\beta-1)/2$. The CVD-$L_{2D}$ relation becomes flatter ($\gamma\to 0$) for thicker clouds. We used this result to constrain the thicknesses of Taurus, Perseus, and Ophiuchus. We conclude that Taurus has a ratio of cloud depth to cloud length smaller than about 1/10-1/8, i.e. it is a sheet. A simple geometric model fit to the linewidth-size relation indicates that the Taurus cloud has a $\sim 0.7$ pc line-of-sight dimension. In contrast, Perseus and Ophiuchus  are thicker and have ratios of cloud depth to cloud length larger than about 1/10-1/8.
\end{abstract}


\section{Introduction}

Stars form in molecular clouds. There is much evidence that molecular clouds are turbulent, e.g. the observed molecular linewidth is much larger than the thermal linewidth. Because the interstellar medium (ISM) is turbulent, its velocity spectrum is self-similar, i.e., a power-law, as evidenced in numerous studies \citep[see the review by][and the references therein]{Hennebelle2012}.

Turbulence in molecular clouds has been studied in a variety of scales and contexts.
\cite{Larson1981} found for the first time a power law relation between the velocity dispersion $\sigma$ and the (projected, 2D) size $L$ of molecular clouds $\sigma_v\propto L^{0.38}$. This power law form relation between the velocity dispersion and the size of molecular clouds is then called Larson's relation. Later observations established the now widely accepted index of 0.5  \citep{Solomon_etal1987,Falgarone1992,Heyer2004}. The former index is close to the power law index $1/3$ of incompressible turbulence, while the latter may indicate density fluctuations within molecular clouds \citep{Vazquez-Semadeni1995}, which is confirmed in observational studies \citep{Turbulence_Brunt2002,Brunt2003} and numerical simulations with increasing sophisticated physics and resolution \citep{Ostriker2001, Offner2008, Kritsuk2013}. However, recent studies on this topic bring more complications. \cite{Zhang2014} find in North American and Pelican molecular clouds a power law index of 0.43 for the Larson's relation, which deviates from the traditional value of 0.5. In another study, \cite{Heyer2009} showed that the normalization of the velocity dispersion-size relation molecular clouds, $v_0=\sigma_v/L^{1/2}$, scales with the surface density of the clouds $\Sigma$ as $v_0\sim \Sigma^{1/2}$, indicating that molecular clouds are in self-gravitational equilibrium.

Traditionally, structure function methods \citep[e.g.][]{Turbulence_Brunt2002} and principal component analysis \cite[PCA,][]{Heyer1997,Brunt2003} are used for characterizing the turbulent velocity field. Recently, \cite{Qian2012} developed the core velocity dispersion (CVD) technique to study turbulence in the Taurus molecular cloud. This method uses the centroid velocity (average velocity of a core weighted with mass) of identified molecular cores to sample the velocity field of spatially and spectrally coherent structures in relatively dense gas. In this paper, we compare the characteristics of the CVD with the gas turbulence and examine whether CVD could trace the general motion  of the underlying gas. \cite{Qian2012} found that the dispersion of the core velocities increases with distance, i.e. CVD$\propto L_{2D}^{0.5}$ between 1 pc and 10 pc, which resembles Larson's relation (hereafter, we will attach the subscript 2D to $L$ as $L_{2D}$ and 3D to $l$ as $l_{3D}$ to emphasize the dimension of the spatial scales). Larson's relation can be explained by the velocity spectrum of a 3D turbulent velocity field \citep{Kritsuk2013}. There is a fundamental difference between the CVD-$L_{2D}$ relation and the turbulent velocity spectrum, in that the turbulent velocity spectrum is three-dimensional, while the spatial scale in the CVD-$L_{2D}$ relation is a projected (2D) scale.

In this work, we expanded upon the previous results and study the influence of the line-of-sight scale (thickness)
on the CVD-$L_{2D}$ relation. We described our observational data in \S\ref{sec:data},  presented our methods in \S\ref{sec:methods}, and discussed results in \S\ref{sec:results}. In \S\ref{sec:conclusion}, we summarized our conclusions and addressed the potential application of the CVD in future studies.

\section{Observational Data}
\label{sec:data}

The $^{13}$CO (J=1-0, 110.2014 GHz) data of Taurus in this work were obtained with the 13.7 m FCRAO telescope between 2003 and 2005. The map is centered at ${\rm RA}(2000.0)=04^h 32^m 44.6^s$, ${\rm Dec}(2000.0)=24^\circ 25' 13.08"$, with an area of $\sim 98\ \rm
deg^2$, and a noise level of 0.1 K  \citep{Narayanan2008}.
The velocity resolution is 0.266 km/s. The $^{13}$CO data of Perseus and Ophiuchus came from the COMPLETE database \citep{Ridge2006}\footnote{http://www.cfa.harvard.edu/COMPLETE/}, which were also observed with the FCRAO telescope. The velocity resolution is 0.066 km/s. The noise levels of Perseus and Ophiuchus are 0.15 K and 0.2 K, respectively.

Following \citet{Qian2012}, we used the GAUSSCLUMPS method in the Starlink software\footnote{http://starlink.jach.hawaii.edu/starlink/} to identify cores from the data cube. The lower thresholds of the peak intensity of cores were set to be 7 times the noise level \citep[see][]{Qian2012}, which were about 0.7 K, 1.0 K, and 1.5 K for Taurus, Perseus, and Ophiuchus, respectively. In Taurus, Perseus, and Ophiuchus, 588, 693, and 270 cores were identified, respectively. The typical size of the cores is around 0.1 pc.

\section{Methods}
\label{sec:methods}

In order to investigate the influence of the cloud thickness on the CVD-$L_{2D}$ relation, we used two approaches.
In the first approach we statistically generate a sample of cores in a velocity field with second order structure function of index 0.5, which corresponds to an energy spectrum of index $\beta=2$. In the second approach we perform hydrodynamic simulations of molecular clouds.

\subsection{Calculation of the CVD}

\begin{figure}[htbp]
\begin{minipage}[b]{0.45\textwidth}
\includegraphics[width=8cm]{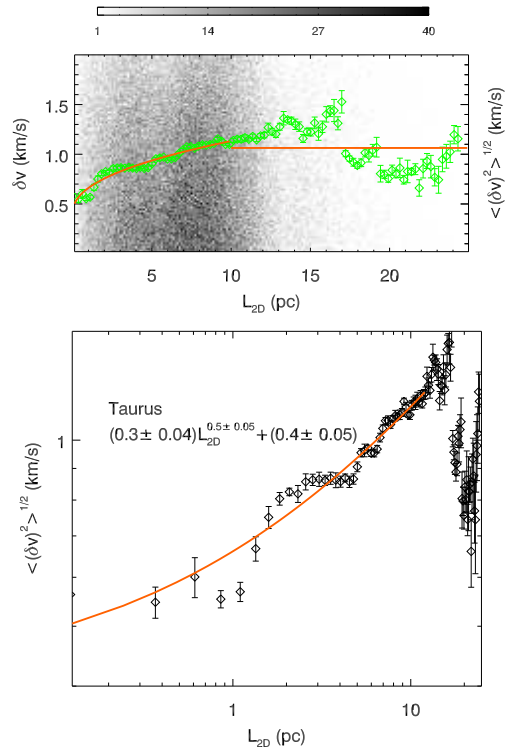}
\end{minipage}
\caption{{\it Top}: Core velocity difference between each core pair, $\delta v$, and the core velocity dispersion
(CVD$\equiv \langle\delta v^2\rangle^{1/2}$) vs.\ the
projected separation, $L_{2D}$, of the cores for the Taurus molecular cloud. The background is an
image of the number of data points in the $\delta v - L_{2D}$ plane, with the
grey scale (density bar at the top of the figure) showing the
density of points. The green diamonds represent the CVD in each
separation bin. The CVD data points with apparent separation $ 0
\leq L_{2D} \leq 10$  pc can be fitted with a power law of CVD
(km/s)$=(0.3\pm 0.04)L_{2D} ({\rm pc})^{0.5\pm 0.05}+(0.4\pm 0.05)$. The horizontal line shows the
mean CVD value, $1.1$ km/s, of data points with $L_{2D}> 10$ pc. {\it Bottom}: The CVD as a function of projected distance. The range of the power law fit is determined in this log-log scale plot. The fitting error is estimated with a bootstrap method.}
\label{fig:vdis_all}
\end{figure}

CVD stands for Core Velocity Dispersion, which is obtained as follows \citep{Qian2012}. First the difference of centroid velocity ($\delta v$) and projected spatial distance ($L_{2D}$) of each pair of cores are calculated. The distribution of $\delta v$ in the $\delta v - L_{2D}$ plane is in grey scale in the top panel of Figure~\ref{fig:vdis_all} for the Taurus molecular cloud. Then we calculate the CVD, which is the mean square root of the values of $\delta v$, i.e. CVD $\equiv <\delta v^2>^{1/2}$,  at different $L_{2D}$ (see the green diamonds in the top panel of Figure~\ref{fig:vdis_all}). Figure~\ref{fig:vdis_all} is similar to Figure 18 in \cite{Qian2012}, but now we focus on the power-law-form CVD-$L_{2D}$ relation between 1 pc and 10 pc, which reflects the pattern of turbulent motion in Taurus.

In order to compare the CVD to a turbulence velocity spectrum, we fit a power law or a broken power law to the CVD-$L_{2D}$ relation in this work. The break point of the broken power law is determined from the log(CVD)-log($L_{2D}$) plot by eye. The fitting error is estimated with a bootstrap method \citep{nr}. We construct 100 samples of core pairs based on the real sample of core pairs, in which every pair in each sample is randomly selected from the data. Then we fit parameters to these 100 samples, and the error is estimated by calculating the root of mean square of the fitted parameters. Since there is degeneracy between the normalization and the index of a power law function, for power law fitting through out this paper, the power law index is first obtained by fitting a linear function to the log-log data and then fixed in the subsequent fitting in linear space, meant to obtain the offset constant.

\subsection{Statistical Core Distribution Model}
\label{subsec:random_cores}

In order to study how well the core velocities correlate with the underlying velocity field and the cloud geometry, we formulate a simple statistical model, the fractional Brownian motion model, of cores within a turbulent molecular cloud. The core sample is generated with procedures similar to those used in generating a turbulent velocity field \citep{Turbulence_Brunt2002,Falgarone2003}. Theoretically, the root of mean square turbulent velocity at scale $l$ (scale length in 3D), $v_l$ , can be determined by the velocity fluctuations over scales smaller than $l$:
\begin{equation}
v_l^2\sim \int^{\infty}_{2\pi/l}E(k)dk\propto \int^{\infty}_{2\pi/l} k^{-\beta}dk\propto l^{\beta-1}\lc
\label{equ:energy_spec}
\end{equation}
where $k$ is the wave number.
If we assume $<v_l^2>^{1/2}\propto l^{\gamma}$, the relation between $\beta$ and $\gamma$ will be $\gamma=(\beta-1)/2$ in 3D. By definition,
\begin{equation}
v_l^2\sim \int v(k)^2 k^2dk \lc
\end{equation}
where $v(k)\propto k^{-\delta}$.

We first generate a 3D velocity field and put roughly 4000 cores on a randomly selected set of grid points. A simple way to do this follows the procedure below. First, we generate a Gaussian random field with dimension 128$\times$128$\times$128 on a 3D grid. Next, we perform a Fourier transform. Then, we process this field in the frequency domain to satisfy the desired power law distribution ($\propto k^{-\delta}$). Finally, we perform an inverse Fourier transformation and normalize the generated field to fulfill the desired variance, which is related to the velocity dispersion.

\subsection{Hydrodynamic Simulations and Synthetic Observations}
\label{subsec:simulation}

 We use hydrodynamic simulations to explore the effects of density/velocity correlations.
We use the {\sc orion} adaptive mesh refinement (AMR) code to perform simulations of turbulent star-forming molecular clouds \citep{truelove98,klein99}.  The simulations use periodic boundary conditions and are meant to represent a piece of a typical low-mass star forming cloud similar to Taurus. The numerical methods we use are similar to previous calculations \citep[e.g.][]{Offner09, Offner13}, so we briefly describe the simulations here and refer the reader to these papers for additional details.

We compare the observations with snapshots from two simulations, each with different turbulent properties (see Table \ref{simprop}). Both simulations have a domain size of 10 pc, contain $\sim 15,000 M_{\odot}$, and have an initial gas temperature of 10 K. Simulation  RT has a turbulent velocity dispersion of 1.52 km s$^{-1}$, such that it satisfies the observed linewidth-size relation at 10 pc, $\sigma_{\rm 2D}= 0.72 R^{0.5}$ km s$^{-1}$ (e.g., \citealt{mckee07}). RT1 and RT2 correspond to two different output times for the RT simulation. They have ages of 0.65 Myr and  1.03 Myr, respectively. The turbulence is initialized by adding random velocity perturbations with an input power spectrum $P(k) \propto k^{0}$ for input wave numbers in the range $k=1\sim 2$. Simulation HD has a velocity dispersion of 1.06 km s$^{-1}$ and the turbulent perturbations have $P(k) \propto k^{-2}$ for $k=1\sim 10$. The latter turbulent perturbations are generated in the same way as those used in section~\ref{subsec:random_cores}. To achieve a turbulent steady-state, in which the power-spectrum and density distribution function are constant in time, we inject the perturbations for two domain crossing times without self-gravity.

After turning on gravity, we continue injecting energy so that the velocity dispersion remains constant. Without continuous energy injection, the turbulence would decay as a result of dissipation in shocks over a crossing time \citep[e.g.][]{Stone1998,MacLow1999}. Once collapse begins, refinement is added to ensure that the Jeans criterion is satisfied for a Jeans number of $N_J=0.125$ \citep{truelove97}. We insert sink particles when the density exceeds the Jeans resolution on the maximum AMR level \citep{krumholz04}. HD adopts an isothermal equation of state, while RT solves the radiative transfer equation in the flux limited diffusion approximation; it also includes radiative feedback from the forming stars \citep{krumholz07,Offner09}. This heating, which is generally limited to regions near sink particles, has minimal impact on the turbulent properties.

RT and HD have $512^3$ and $256^3$ base grids with 2 and 4 AMR levels of refinement, respectively. Here, we only use the data on the base grid, which is comparable to the observational pixel resolution, to produce the synthetic observations (see below). Figure \ref{powersim} shows the velocity energy spectra for the three snapshots. For low wave numbers ($k=1\sim 4$) the energy spectrum is flatter than $k^{-2}$. It steepens above $k\sim 20$ where dissipation occurs. The dissipation scale is set by the numerical grid size and typically occurs on a scale of a few grid cells.

\begin{figure}[htbp]
\centering
\begin{tabular}{c}
\includegraphics[width=8cm]{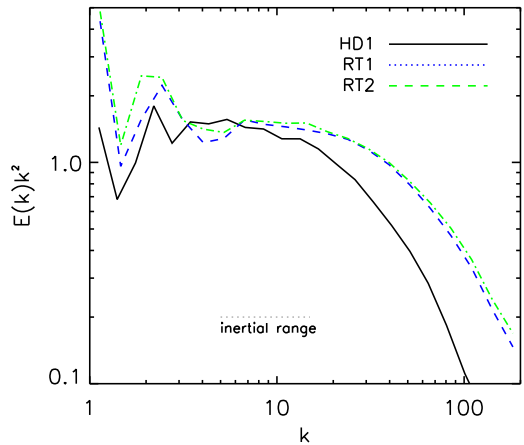}\\
\end{tabular}
\caption{ Simulated velocity energy spectra compensated by $k^2$ for the hydrodynamic simulations. The grey dotted line indicates the inertial range. \label{powersim}}
\end{figure}

In order to compare the simulations directly with the Taurus observations we post-process the outputs with the radiative transfer code {\sc radmc-3d}\footnote{http://www.ita.uni-heidelberg.de/~dullemond/software/radmc-3d/}. We use the Large Velocity Gradient (LVG) approach, which computes the molecule level populations by solving for radiative statistical equilibrium \citep{shetty11}. The input densities, temperatures, and velocities are supplied by the hydrodynamic simulations.

We model unresolved turbulent broadening by adopting a constant micro-turbulence parameter of 0.1 km s$^{-1}$. The H$_2$ number density is defined as $n_{{\rm H}_2}=\rho/(2.8 m_{\rm p})$, where $\rho$ is the gas density, the 2.8 factor accounts for the molecular weight of Helium.  We assume a constant $^{13}$CO abundance of [$^{13}$CO/H$_2$]=$2.5 \times 10^{-6}$ \citep{Frerking1982,Langer1993}, and we adopt the molecular collisional coefficients from \citet{schoier05}. We compute the line emission for the velocities within $\pm5$ km s$^{-1}$ of the line rest frequency and use a channel resolution of $\Delta v=0.039$ km s$^{-1}$ .

To account for differences in resolution and sensitivity, we convolve the synthetic line data with a 45" Gaussian beam, assuming that the simulated cloud lies at a distance of 140 pc. We re-bin the emission such that the pixel size is 20" and the velocity channel width is 0.26 km s$^{-1}$. Finally, we add Gaussian random noise with a standard deviation of $0.3$ K.

In order to assess the effects of cloud thickness on the observations, we perform the line transfer on input volumes of different extents along the line of sight. We compare the cases for line-of-sight thicknesses $L$, $\frac{1}{2}L$, $\frac{1}{4}L$, $\frac{1}{8}L$, and $\frac{1}{16}L$, where $L=10$ pc.  Table \ref{radprop} summarizes all the {\sc radmc-3d} runs.

\begin{deluxetable*}{lcccccc}
\tablecolumns{7}
\tablecaption{Simulation Properties \label{simprop}}
\tablehead{ \colhead{Model\tablenotemark{a}} &
  \colhead{$M$($\msun$)} &
  \colhead{$\mathcal{M}$} &
  \colhead{$L$ (pc)} &
  \colhead{$t_f$(Myr)} &
  \colhead{$N^3$} &
  \colhead{$\Delta x_{\rm min}$(pc)}}
\startdata
RT1  & $1.475\times 10^4$ & 14.0 &  10.0   & 0.65 &  $512^3$    & 0.005        \\ 
RT2  & $1.475\times 10^4$ & 14.0 &  10.0   & 1.03 &  $512^3$    & 0.005        \\ 
HD1    & $1.475\times 10^4$ & 10.0 &  10.0    & 1.06 &  $256^3$    & 0.0025        
\enddata
\tablenotetext{a}{ Run name, total gas mass, domain length, analysis output time, base-grid size, and minimum cell size. RT1 and RT2 are two different output times for the same simulation. They have random perturbations in the range $k=1\sim 2$ with $P(k_{1-2}) \propto k^0$. HD1 has random perturbations $k=1\sim 10$ with $P(k_{1\sim 10})) \propto k^{-2}$. All runs are first evolved for two crossing times without self-gravity and have an initial temperature of 10 K. RT1 and RT2 also include radiative feedback from forming stars (Offner et al. 2009). }
\end{deluxetable*}

\begin{deluxetable*}{lccc}
\tablecolumns{4}
\tablecaption{{\sc radmc} Runs \label{radprop}}
\tablehead{ \colhead{Model\tablenotemark{a}} &
\colhead{$N_x\times N_y\times N_z$} &
\colhead{View} &
 \colhead{($L_x$(pc), $L_y$(pc), $L_z$(pc))}}
\startdata
RT1\_L10z\_M14     &   $512\times512\times 512$ & $z$ & (10, 10, 10)  \\
RT1\_L5z\_M14     &   $512\times512\times 256$ & $z$ & (10, 10, 5)  \\
RT1\_L2.5z\_M14     &   $512\times512\times 128$ & $z$ & (10, 10, 2.5)  \\
RT1\_L0.125z\_M14     &   $512\times512\times 64$ & $z$ & (10, 10, 0.125)  \\
RT1\_L0.0625z\_M14     &   $512\times512\times 32$ & $z$ & (10, 10, 0.0625)  \\
RT1\_L10y\_M14     &   $512\times512\times 512$ & $y$ & (10, 10, 10)  \\
RT1\_L5y\_M14     &   $512\times512\times 256$ & $y$ & (10, 10, 5)  \\
RT1\_L2.5y\_M14     &   $512\times512\times 128$ & $y$ & (10, 10, 2.5)  \\
RT2\_L10y\_M14     &   $512\times512\times 512$ & $y$ & (10, 10, 10)  \\
HD1\_L10z\_M10     &   $256\times256\times 256$ & $z$ & (10, 10, 10)  \\
HD1\_L5z\_M10     &   $256\times256\times 128$ & $z$ & (10, 10, 5)
\enddata
\tablenotetext{a}{ Model name, input {\sc radmc} grid size, line-of-sight view, and domain size. We adopt a constant micro-turbulence value of 0.1 km/s and a doppler catching parameter of 0.25 for all runs. }
\end{deluxetable*}

\section{Results}
\label{sec:results}

\subsection{The CVD of Taurus, Perseus, and Ophiuchus}
\label{subsec:cvd_clouds}

The observed Taurus CVD  for cores identified in $^{13}$CO(1-0) is shown in Figure~\ref{fig:vdis_all}. In the scale range $L_{2D}<10$ pc, the CVD-$L_{2D}$ relation is well-fit by a power law given by CVD$=(0.3\pm 0.04)L_{2D} ({\rm pc})^{0.5\pm 0.05}+(0.4\pm 0.05)$ km s$^{-1}$. The characteristic scale of 10 pc can be clearly seen from the log scale plot in the bottom panel of Figure~\ref{fig:vdis_all}.

\begin{figure}[htbp]
\centering
\begin{tabular}{c}
\includegraphics[width=8cm]{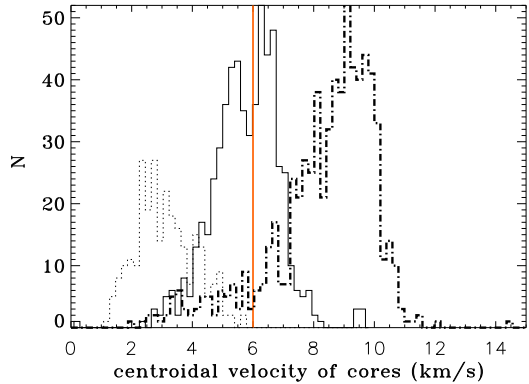}\\
\end{tabular}
\caption{{\it Solid line}:Histogram of the central velocities of cores in the Taurus molecular cloud. There are clearly two groups of cores with different central velocities, divided by the solid line at 6 km/s. {\it Dash-dotted line}: Histogram of the central velocities of cores in Perseus. {\it Dotted line}: Histogram of the central velocities of cores in Ophiuchus.}\label{histvelocity_all}
\end{figure}

\begin{figure}[htbp]
\centering
\begin{tabular}{c}
\includegraphics[width=8cm]{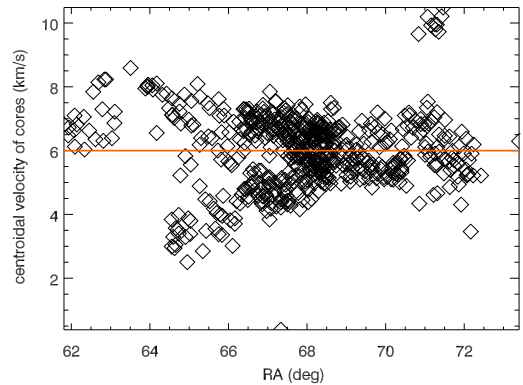}\\
\end{tabular}
\caption{The centroid velocity of cores vs. RA. relation in the Taurus molecular cloud. There are two apparent core groups with the
dividing line roughly lying at 6 km/s (solid line). One group of cores has a velocity gradient.} \label{fig:taurus_core_gradient}
\end{figure}

The solid line in Figure~\ref{histvelocity_all} shows the histogram of the core centroid velocities. The distribution is double-peaked, which could indicate two core populations. This might occur if the Taurus has distinct sub-regions with different line-of-sight velocities. One group has centroid velocities larger then 6 km/s and the other has centroid velocities smaller than 6 km/s. The former group of cores does not have a clear velocity gradient, but the latter group does ($\nabla v\simeq 1$ km/s/deg). A plot of the core centroid velocity versus right ascension as shown in Figure~\ref{fig:taurus_core_gradient} also indicates that there are two core populations.

\begin{figure}[htbp]
\begin{minipage}[b]{0.45\textwidth}
  \includegraphics[width=9.2cm]{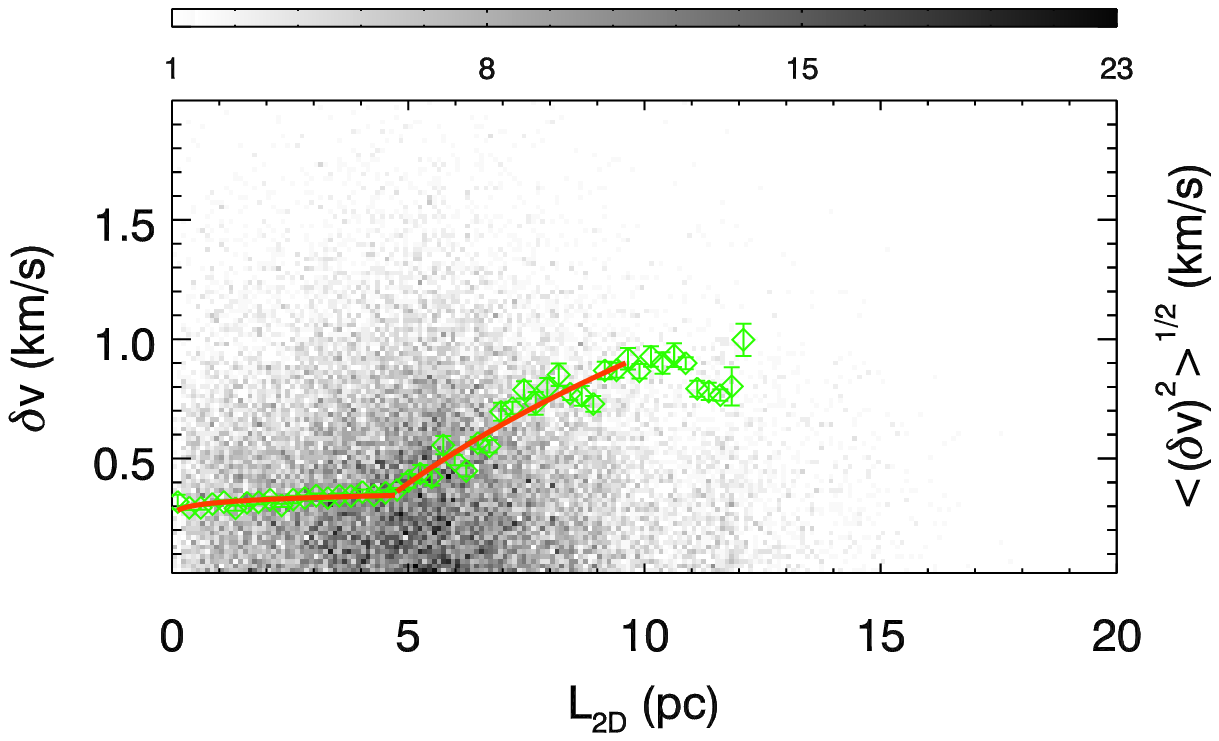}\\
  \includegraphics[width=9.2cm]{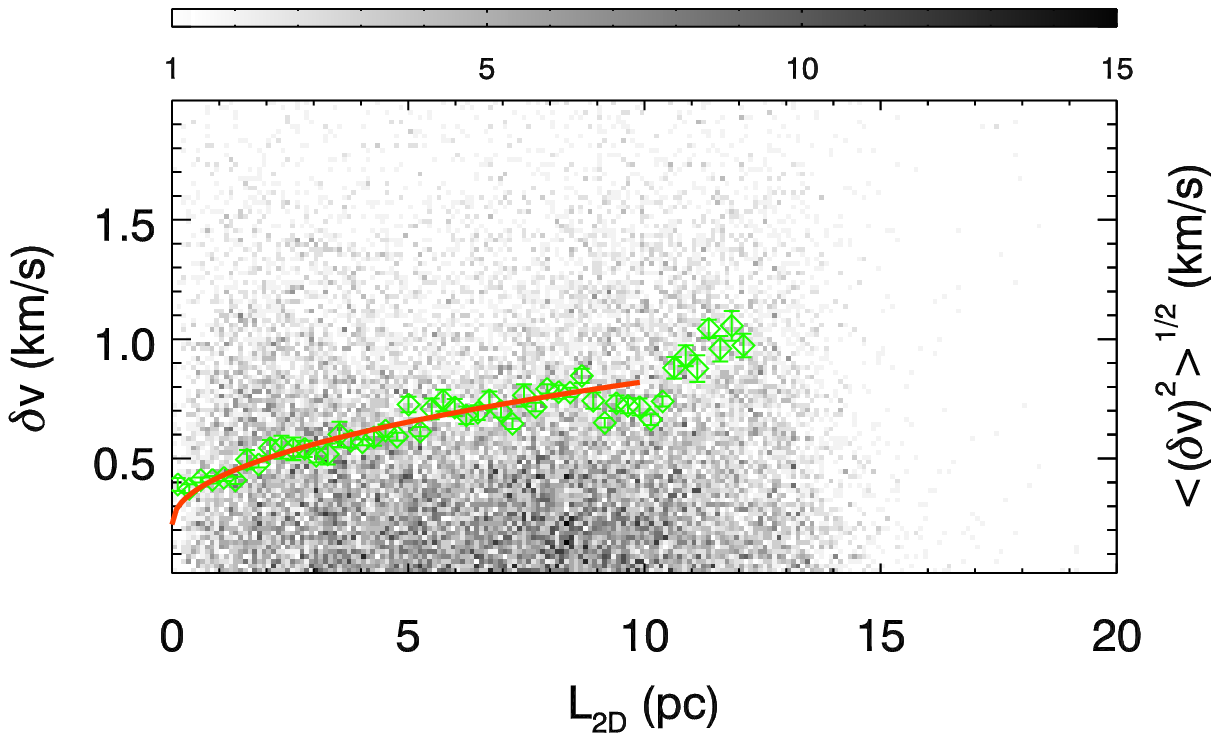}
\end{minipage}
\caption{The CVD plot of the two group of cores in Figure \ref{fig:taurus_core_gradient}: {\it top}: For the case with core centroid velocity greater than 6 km/s, two power laws are needed to fit the data, the power law indices are $0.1\pm 0.03$ ($L_{2D}<5$ pc) and $0.3\pm 0.02$ (5 pc$<L_{2D}<10$ pc), respectively;{\it bottom}: the case with core centroid velocity lower than 6 km/s, the fitted power law index of the CVD-$L_{2D}$ relation is 0.5$\pm$0.3.}\label{fig:taurus_group}
\end{figure}

We performed the CVD analysis of each group of cores and found that the CVD-$L_{2D}$ relation for cores with centroid velocities larger than 6 km/s can be fitted by a broken power law with power law indices $0.1\pm 0.03$ ($L_{2D}<5$ pc) and $0.3\pm 0.02$ (5 pc$<L_{2D}<10$ pc).
Cores with centroid velocities smaller than 6 km/s can be fitted with an index of 0.5$\pm$0.3 (Figure~\ref{fig:taurus_group}). We fit and then subtract the velocity gradient 1 km/s/deg in the latter case and find that the resultant power law index is also 0.5.

\begin{figure}[htbp]
\centering
\begin{tabular}{c}
\includegraphics[width=8cm]{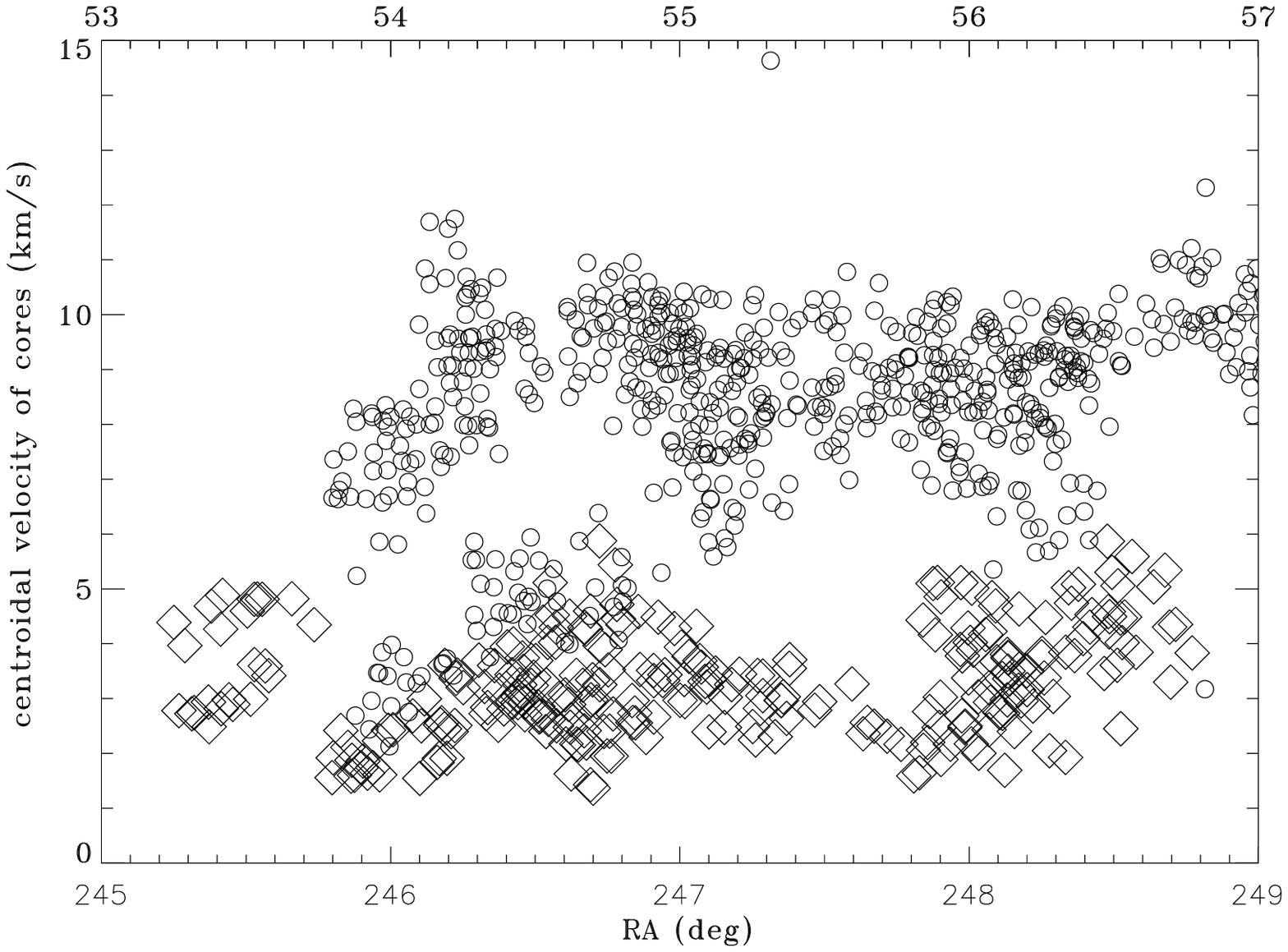}
\end{tabular}
\caption{The centroid velocity of cores vs. RA. in Perseus (diamond, with the top x axis) and Ophiuchus (circle, with the bottom x axis).}\label{fig:perseus_core_gradient}
\end{figure}

\begin{figure}[htbp]
\begin{minipage}[b]{0.45\textwidth}
  \includegraphics[width=9.2cm]{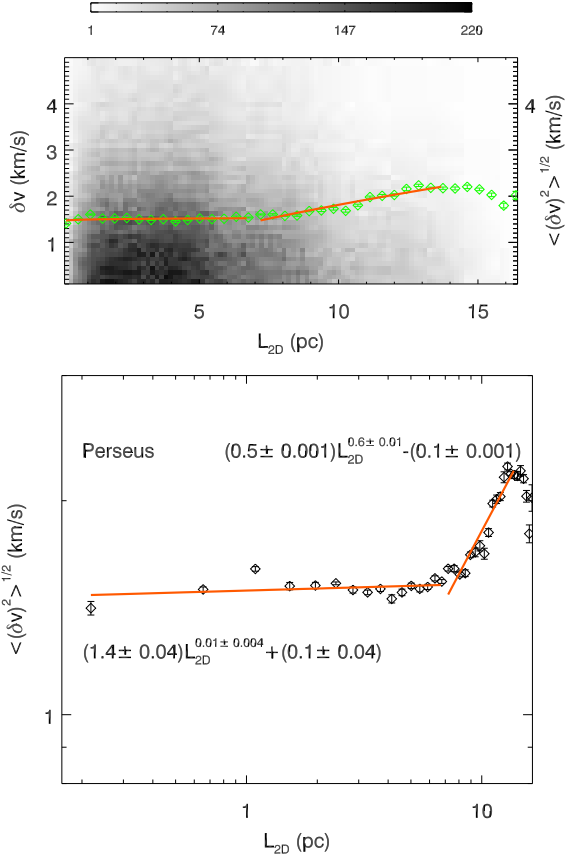}
\end{minipage}
\caption{Similar plot to Figure~\ref{fig:vdis_all}, for Perseus molecular cloud. The CVD data points with apparent separation $0\leq L_{2D} \leq 7$  pc can be fitted with a power law of CVD(km/s)$=(1.4\pm 0.04)L_{2D} ({\rm pc})^{0.01\pm 0.004}+(0.1\pm 0.04)$. While those points in $7\leq L_{2D} \leq 13$  pc can be fitted with CVD(km/s)$=(0.5\pm 0.001)L_{2D} ({\rm pc})^{0.6\pm 0.01}-(0.1\pm 0.001)$. The break point (7 pc) is determined from the lower panel by eye.}\label{fig:pera_vdis_all}
\end{figure}

\begin{figure}[htbp]
\begin{minipage}[b]{0.45\textwidth}
  \includegraphics[width=9.2cm]{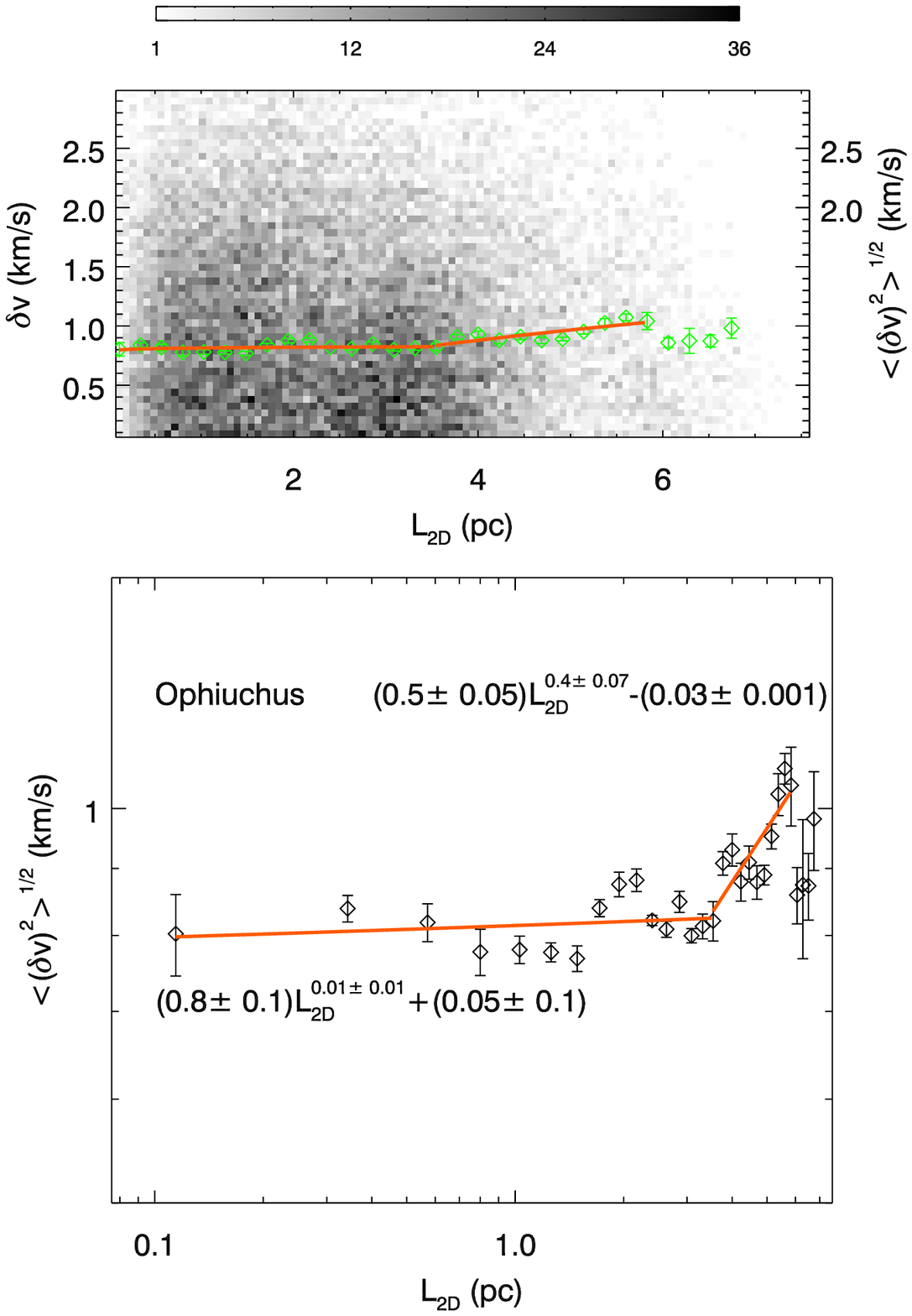}
\end{minipage}
\caption{Similar plot to Figure~\ref{fig:vdis_all}, for Ophiuchus. The CVD data points with apparent separation $0\leq L_{2D} \leq 3.5$  pc can be fitted with a power law of CVD(km/s)$=(0.8\pm 0.1)L_{2D} ({\rm pc})^{0.01\pm 0.01}+(0.05\pm 0.1)$. Those points in $3.5\leq L_{2D} \leq 5$  pc can be fitted with CVD(km/s)$=(0.5\pm 0.05)L_{2D} ({\rm pc})^{0.4\pm 0.07}-(0.03\pm 0.001)$. The break point (3.5 pc) is determined from the lower panel by eye.}\label{fig:oph_vdis_all}
\end{figure}

We have performed a similar analysis for the Perseus and Ophiuchus molecular clouds. In Perseus, the cores do not divide into groups (dash-dotted line in Figure~\ref{histvelocity_all}), although there is a velocity gradient of 4 km/s/deg among the cores (see Figure \ref{fig:perseus_core_gradient}). We perform the fit again after subtracting the gradient and verify that the gradient does not affect our results. The CVD between $0\ {\rm pc}\leq L_{2D} \leq 7$ pc is nearly constant, about 1.5 km/s, while the upturn between $7\ {\rm pc}\leq L_{2D} \leq 13$ pc can be fitted with CVD(km/s)$=(0.3\pm 0.001)L_{2D} ({\rm pc})^{0.6\pm 0.01}$ (Figure~\ref{fig:pera_vdis_all}). The break point (7 pc) is determined from the lower panel by eye. In Ophiuchus, the core velocity distribution is also centrally peaked (dotted line in Figure~\ref{histvelocity_all}). There is no core velocity gradient (Figure~\ref{fig:perseus_core_gradient}), and we find that the  CVD between $0\ {\rm pc}\leq L_{2D} \leq 3.5$ pc is nearly constant, about 0.8 km/s. The upturn in the range of $3.5\leq L_{2D} \leq 5$  pc can be fitted with CVD(km/s)$=(0.5\pm 0.05)L_{2D} ({\rm pc})^{0.4\pm 0.07}-(0.03\pm 0.001)$ (Figure~\ref{fig:oph_vdis_all}). The break point (3.5 pc) is determined from the lower panel by eye.

The breaks in Perseus (7 pc) and in Ophiuchus (3.5 pc) may indicate the thicknesses of the clouds, {\bf which are discussed in section \ref{sec:geometry}}.

\subsection{Influence of geometry}
\label{sec:geometry}

Previous studies of the Taurus molecular cloud found that the CVD follows a power law in the projected length scale (2D distance, $L_{2D}$) with a power law index of 0.5 \citep{Qian2012}. It resembles the Larson's relation. The main difference between these two relations is that the Larson's relation is based on the integrated properties of different clouds, while the centroid velocity of the cores used in the CVD calculation probes the systemic velocities of spatially and spectrally 'coherent' clumps of gas.

In principle, if a cloud has an infinite thickness (but finite optical depth), the CVD-$L_{2D}$ will not hold. The only reason that there is a power law relation between the CVD and the projected length scale $L_{2D}$ is that the cores are distributed in a limited range along the line of sight direction ('thickness'), in which case 2D still bears resemblance to 3D. The difference between 2D and 3D description of turbulence have been studied in some works based on the line centroid velocity analysis \citep[e.g.][]{Falgarone2003,Brunt2004}. In the case where the line of sight scale is much smaller than the transverse scale, {\bf the exponent $\gamma_{2D}$ of the relation between centroid velocity dispersion ($\sigma_v$) and projected scale $L_{2D}$} is related to the exponent of the second order structure function $\gamma_{3D}$ by $\gamma_{2D}=\gamma_{3D}$ \citep{Falgarone2003}. Here, $\gamma_{2D/3D}$ is the power law index of the $\sigma_v$-$L_{2D/3D}$ relation (section \ref{subsec:random_cores}). However, in general, the density inhomogeneity will produce $\gamma_{2D}\approx\gamma_{3D}\approx 0.5$. When the effect of the density inhomogeneity is diminished in the late time of decaying turbulence, the relation between the 2D power law index $\gamma_{2D}$ and the 3D one $\gamma_{3D}$ is $\gamma_{2D}\approx \gamma_{3D}+0.5$ \citep{Brunt2004}. Thus it is theoretically expected that the depth of the cloud would affect the relation between velocity fluctuations (e.g., in the CVD) and projected scale. We explore this effect here using the simulations described in sections \ref{subsec:random_cores} and \ref{subsec:simulation}.

\begin{figure}[htbp]
\includegraphics[width=8cm]{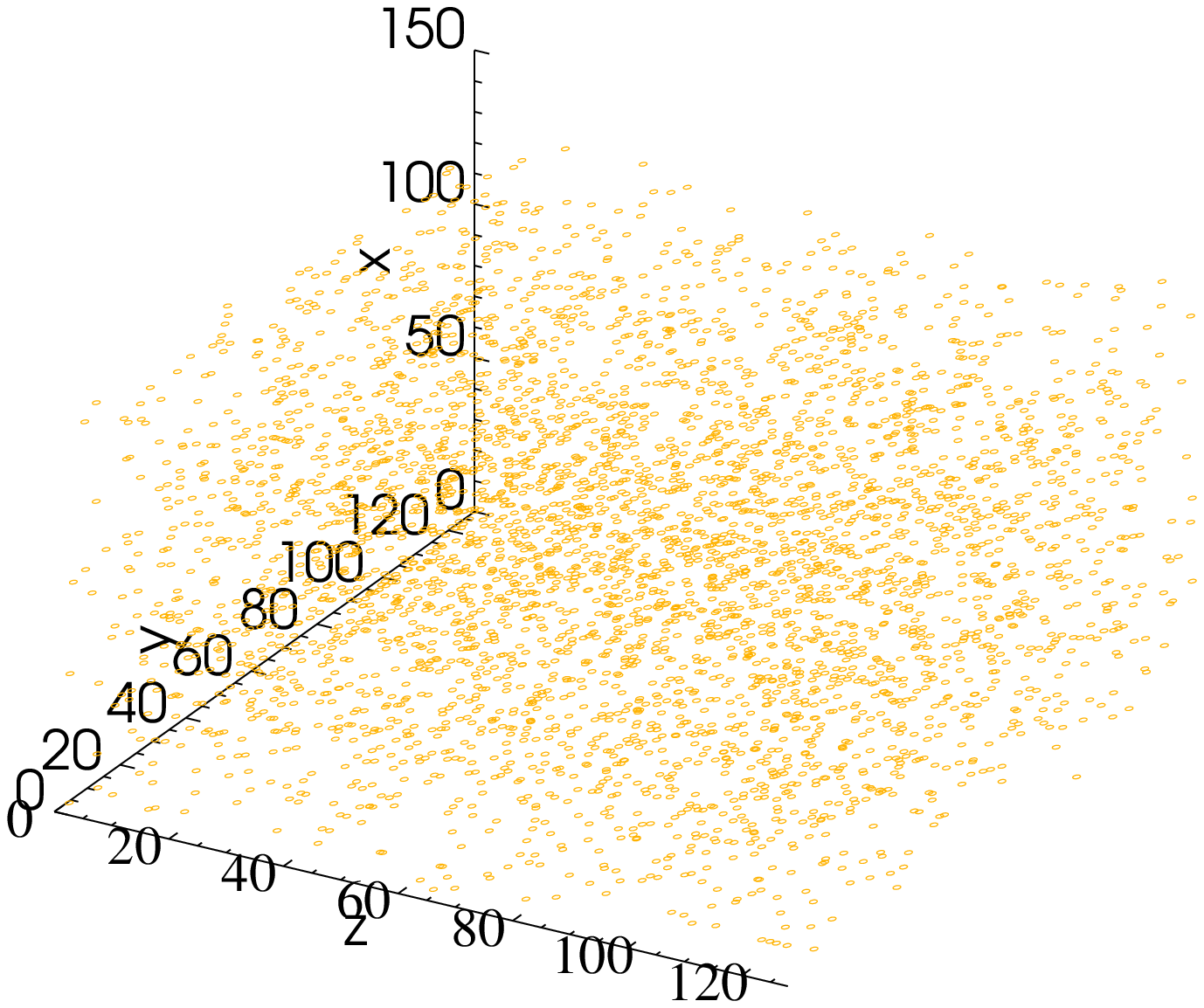} \includegraphics[width=7cm]{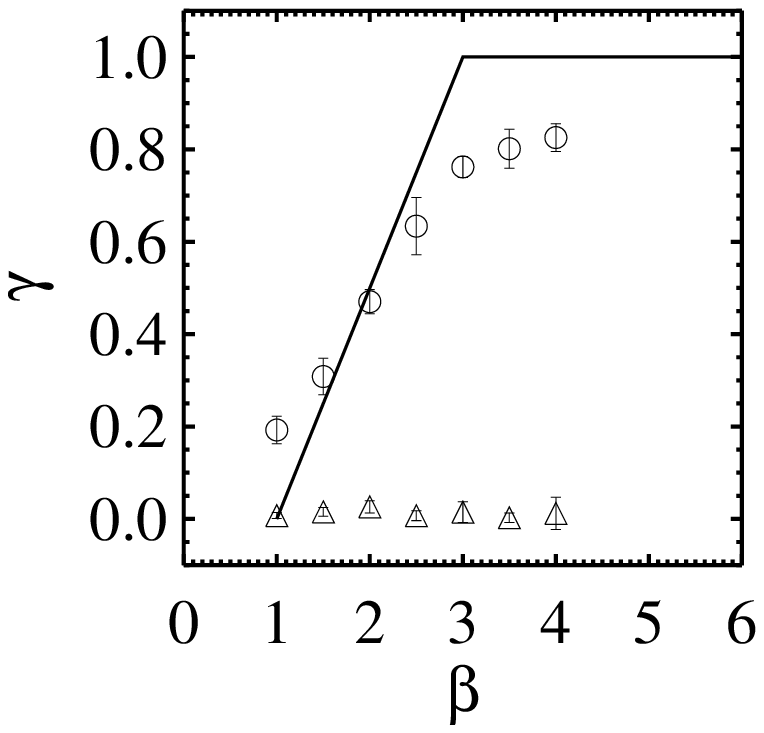}
\caption{ {\it Top:} Distribution of cores in a $128\times 128\times 128$ grid. Each dot represents a core. {\it Bottom:} Relation between the power law index of energy spectrum ($\beta$, see equation \ref{equ:energy_spec}) and the power law index of the CVD-$L_{2D}$ ($l_{3D}$) relation ($\gamma$). The $1< \beta <3$ part of the solid line shows the theoretical relation between $\beta$ and $\gamma$, $\gamma=(\beta-1)/2$. For $\beta>3$, $\gamma$ asymptotically approaches unity. The circles (3D) and triangles (2D) are the results of the model.
\label{core128}}
\end{figure}

As a limiting case of large thickness, in the statistical (fractional Brownian motion) modelling mentioned in section \ref{subsec:random_cores}, we used a $128\times 128\times 128$ grid (with Relative Thickness $\mathcal{R}\equiv z/x=1$) to generate the velocity field. We distribute 4000 cores randomly on the grid points (top panel of Figure \ref{core128}). Each core is assigned a centroid velocity  corresponding to the velocity at its grid location. Thus, we assume each core perfectly traces the underlying velocity field. The power law index of the core velocities is then fitted as shown in Figure~\ref{core128}. The distribution of cores in the $128\times 128\times 128$ grid is shown in the top panel of Figure~\ref{core128}. Each dot represents a core. The bottom panel of Figure~\ref{core128} shows the plots of the power law index of the CVD-$L_{2D}(l_{3D})$ relation ($\gamma$) versus the power law index of the energy spectrum ($\beta$) of the velocity field. Circles are results from the CVD-$l_{3D}$ relation, while triangles are from the CVD-$L_{2D}$ relation. For $1< \beta <3$ the relation between $\gamma$ and $\beta$ of the 3D scenario is close to the theoretical relation between $\beta$ and $\gamma$, $\gamma=(\beta-1)/2$, however, the 2D scenario is quite different, with the CVD-$L_{2D}$ slope being consistent with zero. This reflects the difference between $L_{2D}$ and $l_{3D}$ when the thickness is large. For $\beta>3$, $\gamma$ asymptotically approaches unity \citep{Turbulence_Brunt2002}.

\setlength{\floatsep}{10pt plus 3pt minus 2pt}

\begin{deluxetable*}{lc}
\tablecolumns{2}
\tablecaption{Summary of data fitting \label{tab:result}}
\tablehead{ \colhead{Model\tablenotemark{a}} &
 \colhead{Power law index}
 }
\startdata
Taurus cores & $0.5\pm 0.05$\\
Taurus cores ($>6$ km/s) & $0.3\pm 0.02$\\
Taurus cores ($\le6$ km/s)& $0.5\pm 0.3$\\
Taurus $^{12}$CO gas   & $0.4\pm 0.01$\\
Taurus $^{13}$CO gas   & $0.5\pm 0.03$, $0.2\pm 0.01$\\
Perseus cores (0 pc$\sim$ 7 pc)  &  $0.01\pm 0.004$\\
Perseus cores (7 pc$\sim$ 13 pc)  &  $0.6\pm 0.01$\\
Ophiuchus cores   & $0.01\pm 0.01$\\
RT1\_L0.0625z\_M14 & $0.2\pm 0.1$ \\
RT1\_L2.5z\_M14    &  $0.05\pm 0.11$ \\
RT1\_L5z\_M14     & $0.05\pm 0.01$\\
RT1\_L10z\_M14     &  $0.1\pm 0.005$\\
HD1\_L10z\_M10      & $0.05\pm 0.02$
\enddata
\tablenotetext{a}{Models from Table 2 are only listed if a sufficient number of cores is identified by GAUSSCLUMPS to perform the analysis.}
\end{deluxetable*}

\begin{figure}[htbp]
\includegraphics[width=8cm]{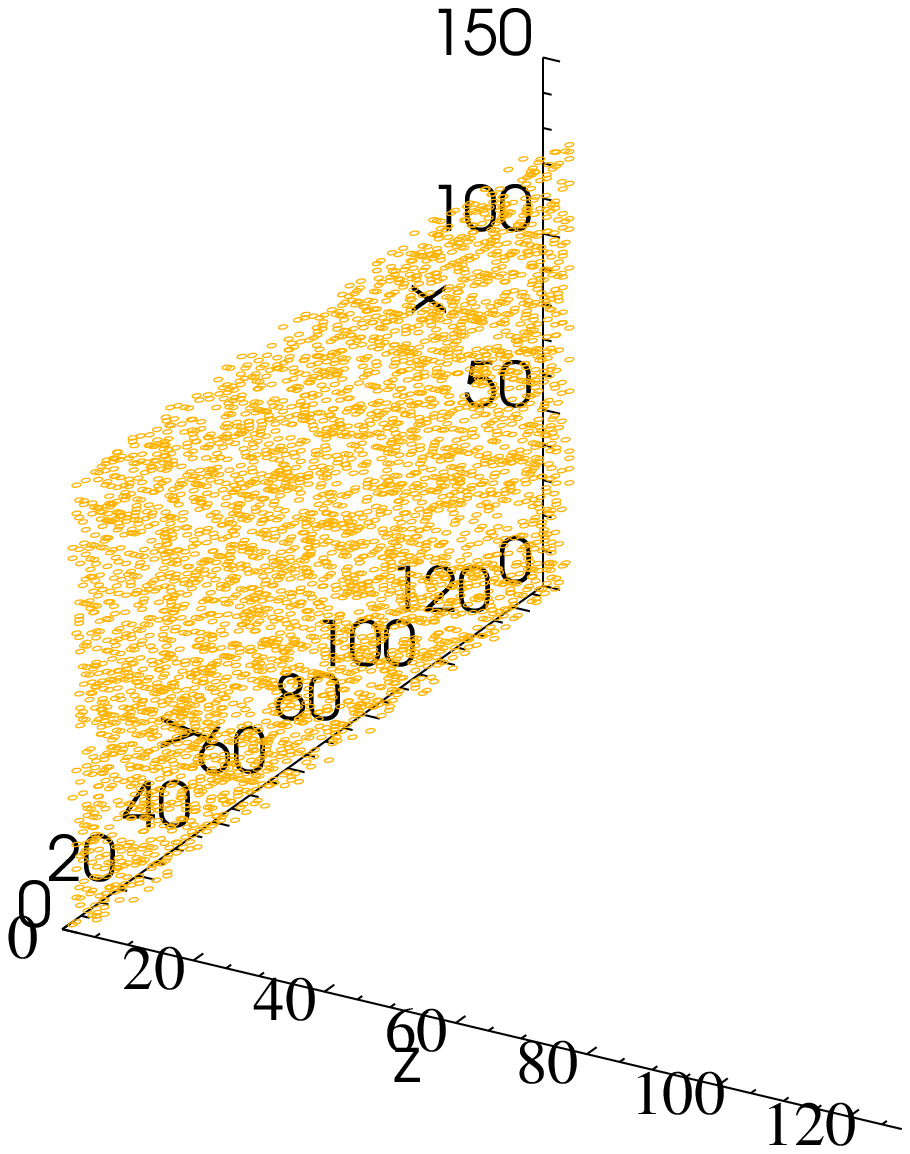} \includegraphics[width=7cm]{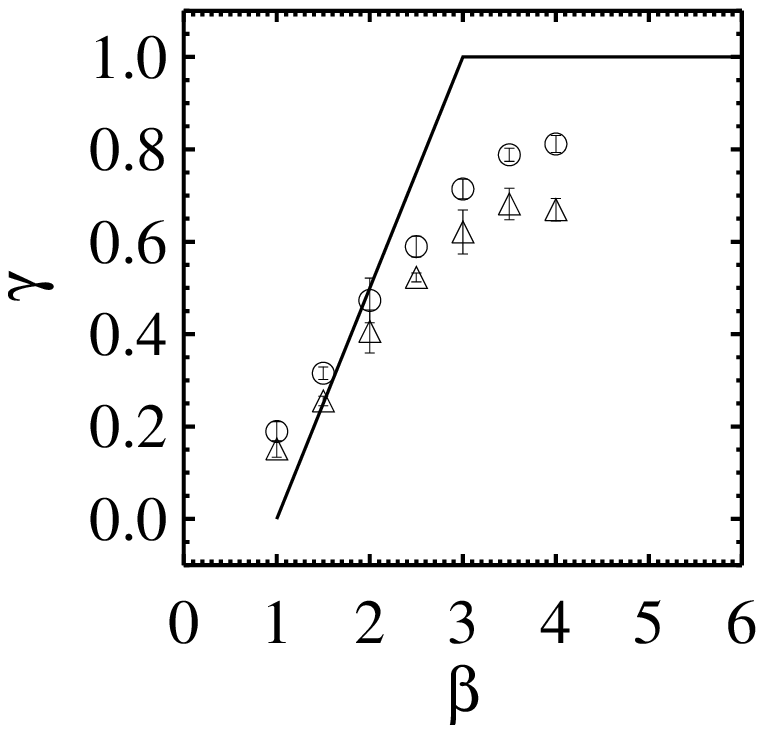}
\caption{ Similar to figure \ref{core128}, for a $128\times 128\times 4$ grid.
\label{core4}}
\end{figure}

As a limiting case of small thickness, we consider a $128\times 128\times 4$ slab (the relative thickness $\mathcal{R}=1/32$) from the previous $128\times 128\times 128$ grid ($\mathcal{R}=1$) and put 4000 cores randomly  on the grid points (Figure \ref{core4}). The power law index of the CVD-$L_{2D}$ ($l_{3D}$) relation is then fitted based on this core sample (Figure~\ref{core4}). The distribution of cores in a $128\times 128\times 4$ grid is shown in Figure~\ref{core4}. The bottom panel of Figure~\ref{core4} shows the power law index of the CVD-$L_{2D}(l_{3D})$ relation ($\gamma$) as a function of the power law index of the energy spectrum ($\beta$) of the velocity field. As in Figure~\ref{core128}, the circles indicate results from the three-dimensional CVD ($l_{3D}$) relation, while triangles are from the projected CVD relation. For $1< \beta <3$ the relation between $\gamma$ and $\beta$ of both 3D and 2D relation is close to the theoretical relation between $\beta$ and $\gamma$, $\gamma=(\beta-1)/2$. For $\beta>3$, $\gamma$ asymptotically approaches unity \citep{Turbulence_Brunt2002}. The 3D and 2D scenarios are now similar. This is natural since the 2D length scale of a thin slab is approximately equal to the 3D length scale.

\begin{figure}[htbp]
\begin{minipage}[b]{0.45\textwidth}
  \includegraphics[width=9.2cm]{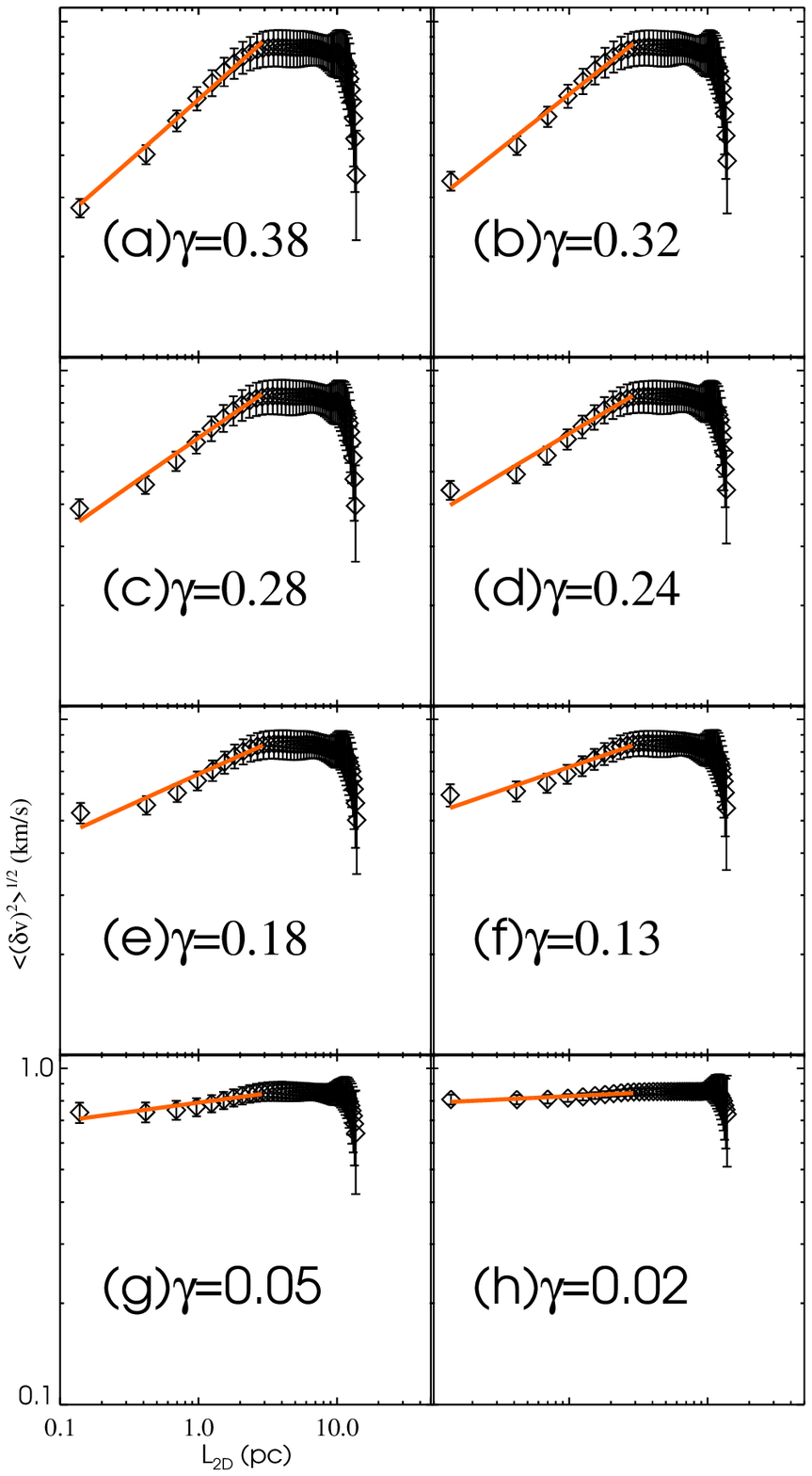}
\end{minipage}
\caption{ CVD-$L_{2D}$ plots similar to Figure~\ref{fig:vdis_all}, for fractional Brownian motion model ($\beta=2$, see equation \ref{equ:energy_spec}) of cores distributed in grids with relative thicknesses: (a)1/32; (b)1/16; (c)1/10; (d)1/8; (e)1/5; (f)1/4; (g)1/2; (h)1. The power law is fitted to the data points with 0.1 pc $<L_{2D}<$ 3 pc. }\label{fig:vdis_fft_errorbar}
\end{figure}

\begin{figure}[htbp]
\centering
\includegraphics[width=8cm]{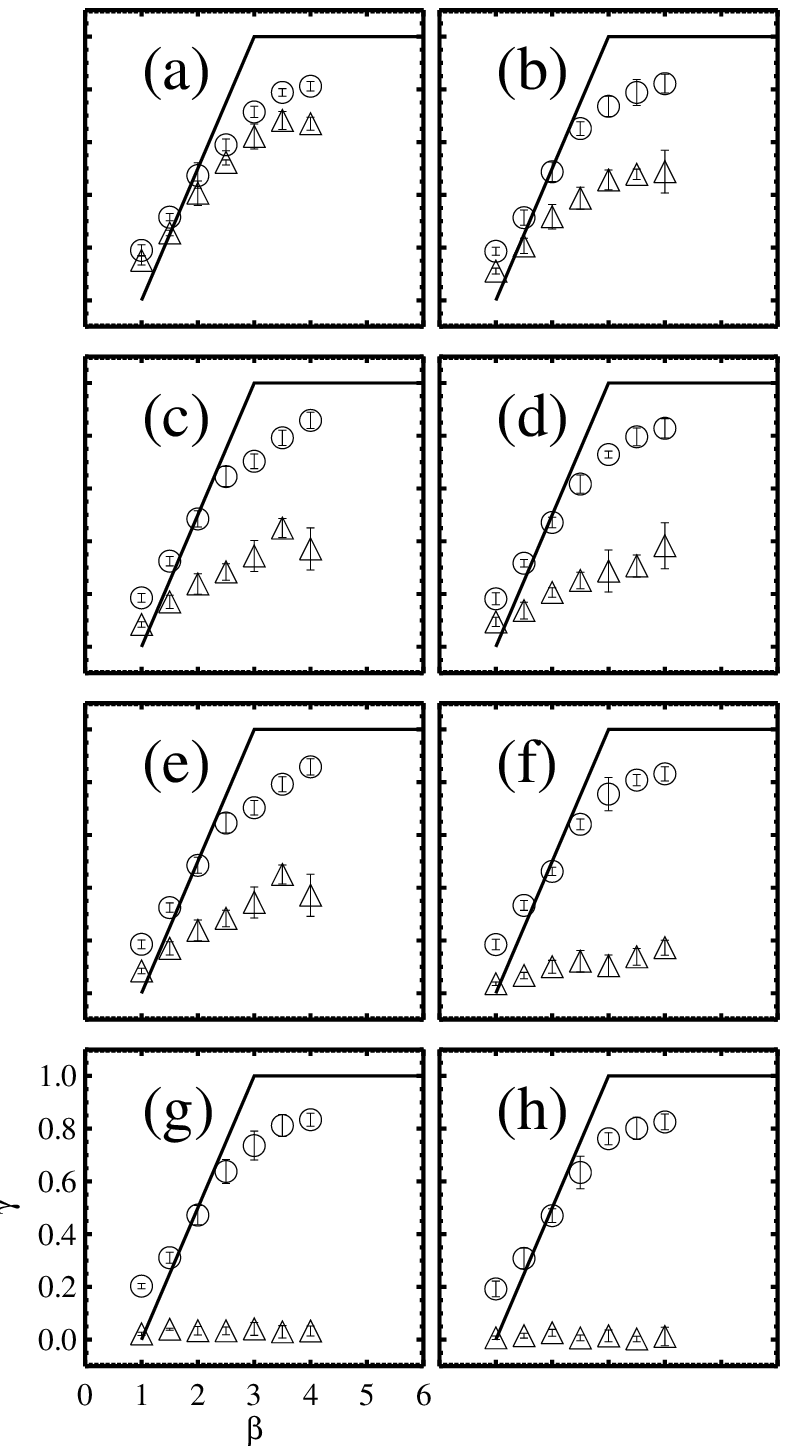}
\caption{Relation between the power law index of energy spectrum ($\beta$) and the power law index of the velocity field ($\gamma$). The $1< \beta <3$ part of the solid line shows the theoretical relation between $\beta$ and $\gamma$, $\gamma=(\beta-1)/2$. The circles (3D) and triangles (2D) are the results of the model. Relative thicknesses in each panel: (a)1/32; (b)1/16; (c)1/10; (d)1/8; (e)1/5; (f)1/4; (g)1/2; (h)1.
\label{compare_4_8_16_32_64_128}}
\end{figure}

{\bf We studied the CVD-$L_{2D}$ relation of the fractional Brownian motion model fields of different relative thicknesses $\mathcal{R}=1/32, 1/16, 1/10, 1/8, 1/5, 1/4, 1/2, 1$. The resulting CVD-$L_{2D}$ relation with $\beta=2$ are shown in Figure~\ref{fig:vdis_fft_errorbar}, with the fits to the relation and resulting $\gamma$ values overlayed.} The panels in Figure~\ref{compare_4_8_16_32_64_128} show the dependence of the $\beta-\gamma$ relation on the thickness of a cloud, where $\mathcal{R}$ is
1/32, 1/16, 1/12, 1/8, 1/5, 1/4, 1/2 and 1. We find that there is a critical
relative thickness of $\mathcal{R}\approx 1/10 - 1/8$, above which the
CVD-$L_{2D}$ relation significantly deviates from the CVD-$l_{3D}$ relation.
This deviation is larger for steeper energy spectra. Since the
CVD-$L_{2D}$ relation for Perseus and Ophiuchus is flat below
7 pc and 3.5 pc, their thickness/transverse extent ratio must be
$\mathcal{R}> 1/10 - 1/8$. With transverse extents of 16 pc and 8 pc for
Perseus and Ophiuchus respectively, this corresponds to a thickness $h>1.6$ pc-2.0 pc for
Perseus and $h>0.8$ pc-1.0 pc for Ophiuchus. Conversely, the CVD-$L_{2D}$
relation for Taurus is consistent with a $\mathcal{R}<1/10-1/8$ or $h<2.5$ pc-
3 pc.

{\bf The upturn of the CVD-$L_{2D}$ relation seen in the observations of Perseus (at 7 pc) and Ophiuchus (at 3.5 pc) are also seen in the fractional Brownian motion simulation (Figure \ref{fig:vdis_fft_errorbar}), albeit less obviously. Nonetheless, the location of the upturn in the fractional Brownian motion simulation seem to correlate with the thicknesses  of the data cubes, which suggests that the differences in the upturn locations in the observations of Perseus and Ophiuchus track the differences in their thicknesses.}

To further constrain the depth of molecular clouds, we have developed a simple geometrical model of how the integrated linewidth of a cloud depends on its transverse scale and thickness. This is discussed in the next section (section \ref{sec:size_linewidth}).

\subsection{Constraining cloud thickness from the linewidth-size relation}
\label{sec:size_linewidth}

\begin{figure}[htbp]
\centering
\begin{tabular}{c}
\includegraphics[width=8cm]{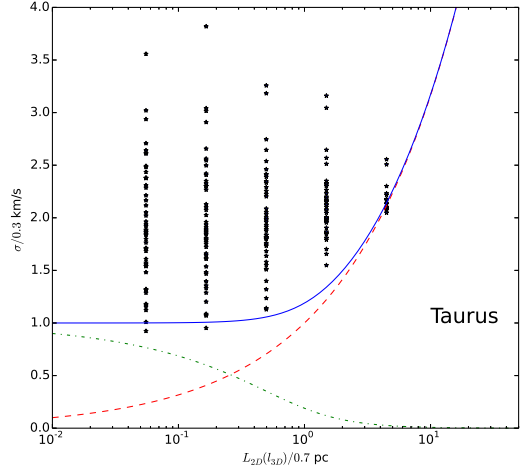}\\
\end{tabular}
\caption{ The solid line denotes the function ($\sigma$/$v_0$)$=\left(\sqrt{(L_{2D}/h)^2+1}\right)^{1/2}$, while the dashed line represents the function ($\sigma$/$v_0$)$= (l_{3D}/h)^{1/2}$. It can be seen that the difference becomes small at large $L_{2D}/h$.
$\sigma$ is normalized with a characteristic velocity $v_0=$0.3 km/s, while $L_{2D}$($l_{3D}$) is normalized with thickness $h=$0.7 pc. The difference between these two function is plotted with dash-dotted line. The asterisks shows the linewidth measured at different scales. The lower envelope of these points probes the thinnest part of the cloud \citep[cf.][]{Li2008}. It flattens at the small scales. \label{width_L}}
\end{figure}

In order to understand this critical value of the relative thickness $\mathcal{R}$, we perform a simple analysis of the relation between the linewidth $\sigma$ and 2D/3D scale $L_{2D}$/$l_{3D}$, and compare with the observational data. For a cloud with an energy spectrum of index $\beta=2$ the integrated line-width is related to the 3D size by
\begin{equation}
\sigma\propto l_{3D}^{0.5},
\label{CVD_3D}
\end{equation}
where $l_{3D}=\sqrt{L_{2D}^2+h^2}$, $h$ is a typical scale along the line of sight of the molecular cloud. So we have
\begin{equation}
\sigma\propto \left(\sqrt{L_{2D}^2+h^2}\right)^{0.5}.
\label{CVD_2D}
\end{equation}
In the limit of $L_{2D}\gg h$, the above two relations have approximately the same function form. In the limit of $L_{2D}\ll h$, $\sigma$ would not depend on $L_{2D}$. The two functions are illustrated in Figure~\ref{width_L}. The specific value of $\mathcal{R}=1/10-1/8$ of the threshold is not clear from this figure, although the general trend that the difference between the $\sigma$-$L_{2D}$ relation and the $\sigma$-$l_{3D}$ relation becomes small at large transverse scales is clear.

We have applied this simple geometric model to Taurus, Perseus, and Ophiuchus. We selected random locations in Taurus and measured the average linewidth in groups of concentric circular regions centered on these locations. This analysis gives a lower limit to the overall thickness of Taurus. The lower envelope of the points on the $\sigma-L_{2D}(l_{3D})$ plane provides a lower limit on the average thickness of the cloud. This lower envelope can be fitted with ($\sigma$/$v_0$)$=\left(\sqrt{(L_{2D}/h)^2+1}\right)^{1/2}$ (Figure~\ref{width_L}), and $v_0=$0.3 km/s, $h=$0.7 pc.

\begin{figure}[htbp]
\centering
\begin{tabular}{c}
\includegraphics[width=8cm]{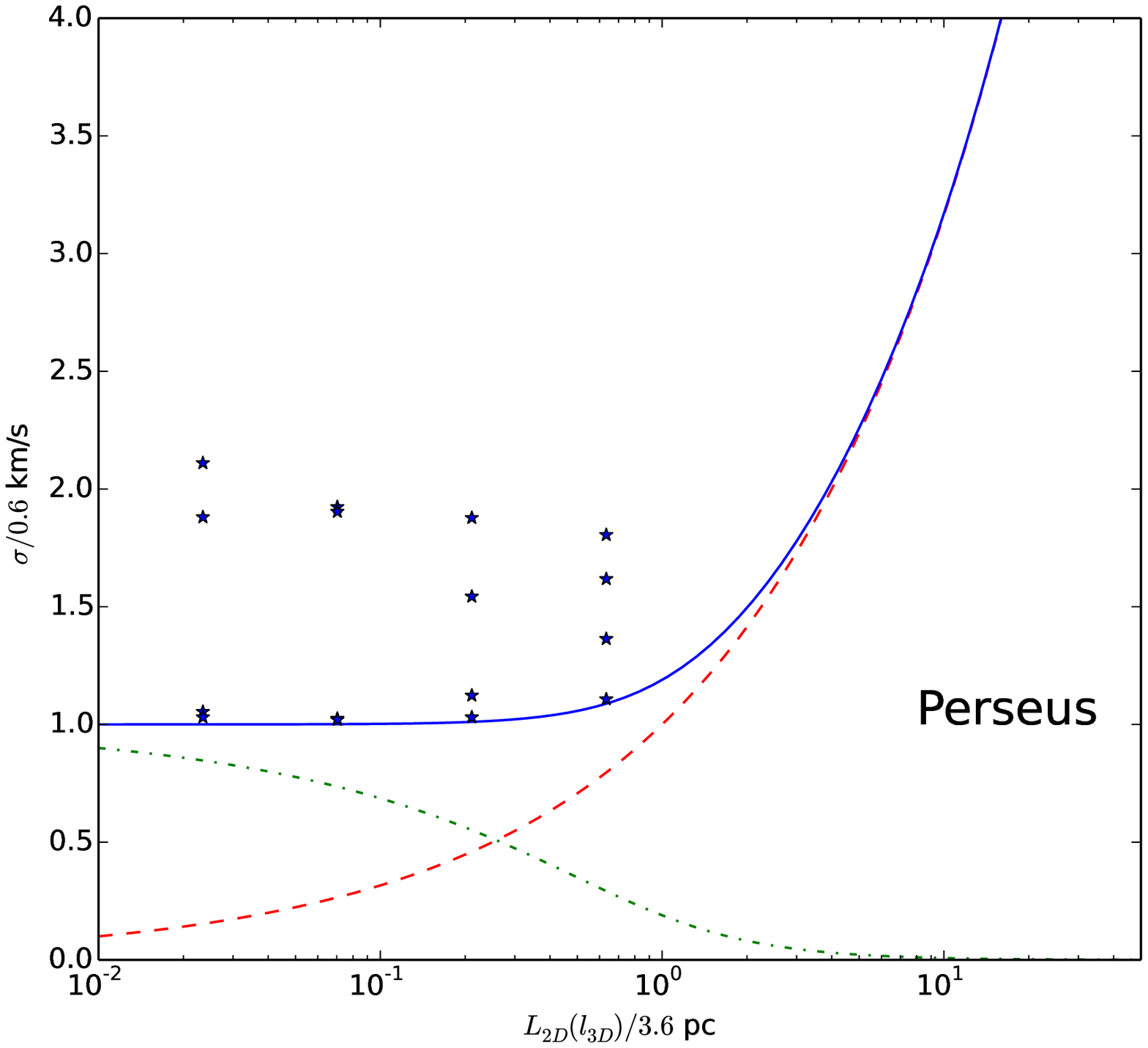}\\
\end{tabular}
\caption{ The same analysis of Perseus molecular cloud as Figure~\ref{width_L}. The characteristic velocity $v_0$ and the thickness $h$ are 0.6 km/s and 3.6 pc, respectively. Since the upturn similar to that in figure \ref{width_L} is not clearly seen, here $h$ is only a lower limit of the thickness of Perseus. \label{pera_width_L}}
\end{figure}

\begin{figure}[htbp]
\centering
\begin{tabular}{c}
\includegraphics[width=8cm]{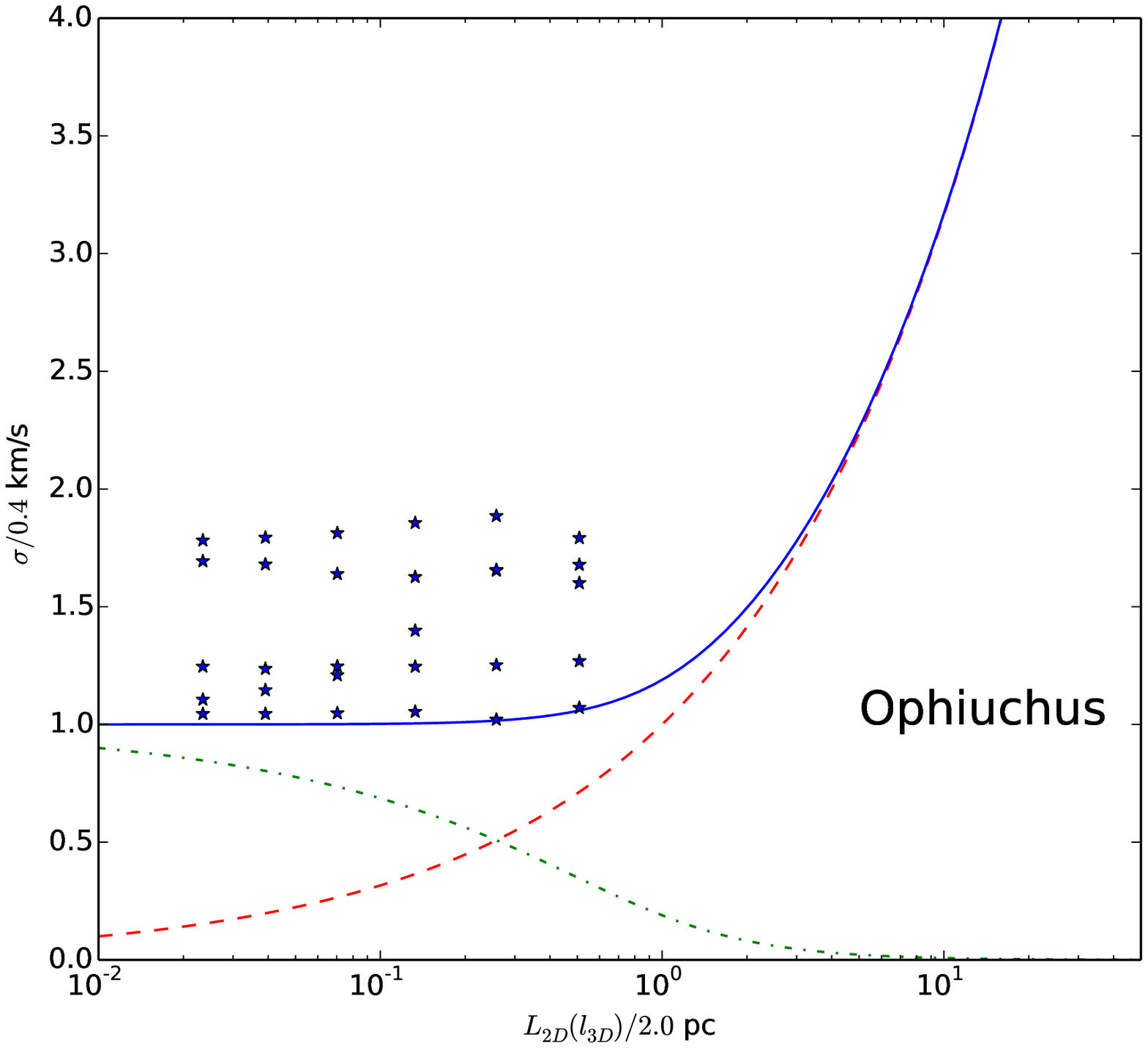}\\
\end{tabular}
\caption{ The same analysis of Ophiuchus molecular cloud as Figure~\ref{width_L}. The characteristic velocity $v_0$ and the thickness $h$ are 0.4 km/s and 2.0 pc, respectively. Since the upturn similar to that in figure \ref{width_L} is not clearly seen, here $h$ is only a lower limit of the thickness of Ophiuchus. \label{oph_width_L}}
\end{figure}

We applied a similar analysis to the Perseus (Figure~\ref{pera_width_L}) and Ophiuchus (Figure~\ref{oph_width_L}) molecular clouds and found that The lower envelope of the points on the $\sigma-L_{2D}(l_{3D})$ plane is nearly flat in both clouds. Note that Perseus and Ophiuchus are elongated, so the spatial dynamic range for concentric circular regions average is smaller that of CVD analysis. The fitted parameters are $v_0=$0.6 km/s, $h=$3.6 pc for Perseus, and $v_0=$0.4 km/s, $h=$2.0 pc for Ophiuchus. Note the thicknesses $h$ of Perseus and Ophiuchus are both lower limits, since the upturn as Taurus is not seen.

In this work, the thickness of a molecular cloud has been estimated with the slope of the CVD-$L_{2D}$ relation, the fitting to the lower envelope of the $\sigma$-$L_{2D}$ relation for all the three clouds, and the upturn location of the CVD-$L_{2D}$ relation for Perseus and Ophiuchus. 1) Based on the slope of the CVD-$L_{2D}$ relation, we conclude that the relative thickness $\mathcal{R}$ of Taurus is smaller than 1/10-1/8, while $\mathcal{R}$ of Perseus and Ophiuchus are larger than 1/10-1/8. The relevant transverse scales are about 25 pc, 16 pc, and 8 pc for Taurus, Perseus, and Ophiuchus, respectively. The estimated upper limit of the thickness of Taurus  is 2.5 pc-3.0 pc. The estimated lower limits of the thickness of Perseus and Ophiuchus are 1.6 pc-2.0 pc and 0.8 pc-1.0 pc, respectively. 2) The characteristic thickness $h$ of Taurus is 0.7 pc from the fitting to the lower envelope of the $\sigma$-$L_{2D}$ relation, which is consistent with the upper limit obtained in 1). The lower limits of the thickness $h$ of Perseus and Ophiuchus are 3.6 pc and 2.0 pc, respectively. 3) The thicknesses possibly indicated by the upturn location of the CVD-$L_{2D}$ relation of Perseus and Ophiuchus are 7 pc and 3.5 pc, respectively. They are consistent with the lower limit obtained in 1) and 2).

The results are consistent with previous studies. For example, \cite{Arce2011} used the morphologies of bubbles within Perseus to conclude that its thickness is about 15\% to 30\% of its transverse scale. This is consistent with our conclusion that the depth of Perseus is larger than 15\% of its projected size. In addition, \citep{Loren1989b} find that the linewidth-size relation may not hold in Ophiuchus, which could partially explain the flat CVD slope we find. \citet{observe_lognormal} find that the average dust extinction of Ophiuchus is much higher than that of other nearby clouds of similar projected size. This can be explained if Ophiuchus is extended along the line-of-sight, and thus, $\mathcal{R}>>1/10-1/8$. The upturning point of the CVD in Figure~\ref{fig:pera_vdis_all} and Figure~\ref{fig:oph_vdis_all} may also be an indicator of the thickness of the clouds. In Figures \ref{width_L}, \ref{pera_width_L} and \ref{oph_width_L}, the linewidth at the flattened part of the linewidth-size relation is 0.3 km/s to 0.6 km/s, which is larger than the thermal linewidth (at 10 K, the thermal linewidth of $^{12}$CO is about 0.07 km/s). The flattening of the linewidth-size relation is little affected by the thermal broadening.

\subsection{Influence of the turbulence}

We used hydrodynamic simulations to study the impact of density-velocity correlations on projection effects \citep[e.g.][]{Brunt2004}. The transverse scale of the simulated clouds is 10 pc, and we varied the cloud thickness from 10 pc to 0.625 pc, by considering thinner sub-sections of the simulated clouds. When the thickness is smaller than 2.5 pc ($\mathcal{R}<1/4$), there are not enough cores in the 3D data cubes, so the errors are larger.

\begin{figure*}[htbp]
\begin{minipage}[b]{\textwidth}
\centering
  \includegraphics[width=16cm]{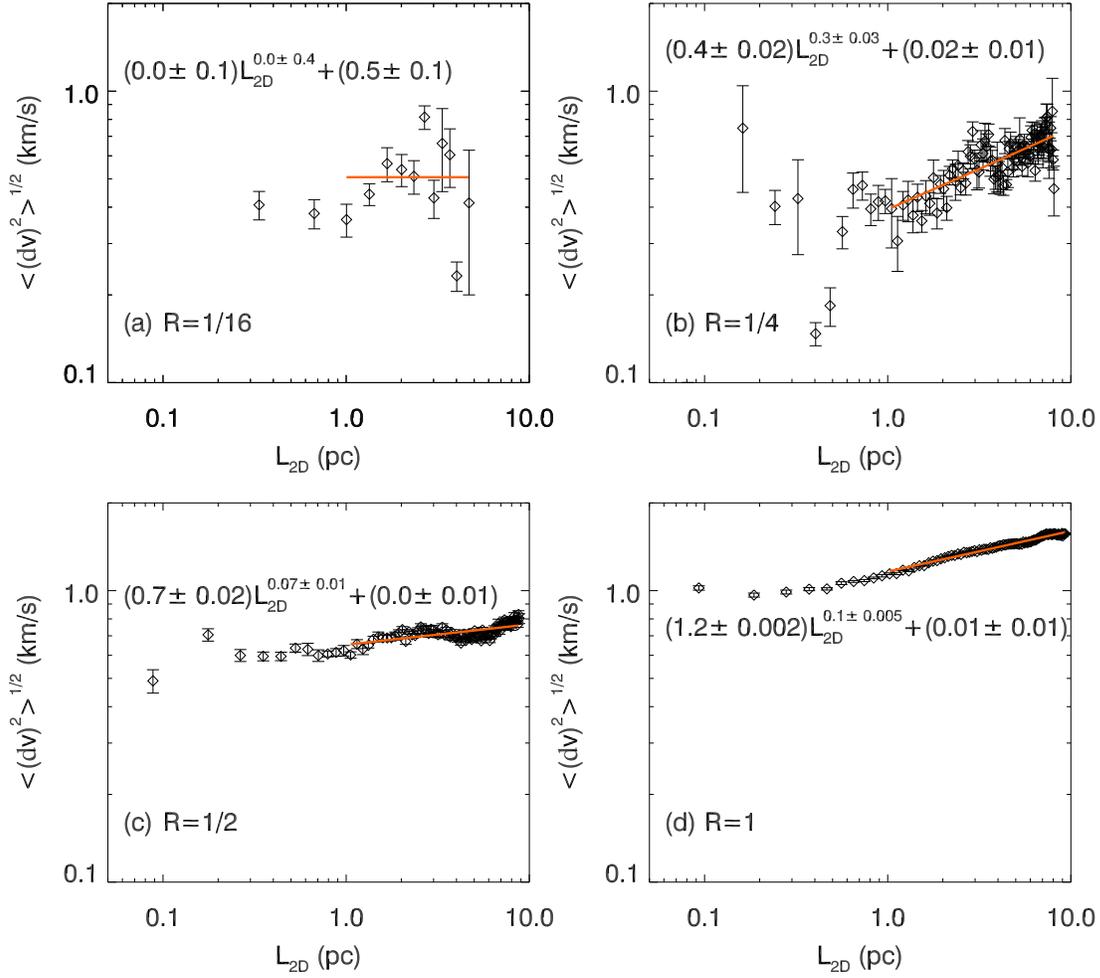}
\end{minipage}
\caption{ The CVD plot for a simulated cloud with a line of sight thickness of (a) 0.625 pc (with $\mathcal{R}=1/16$);(b) 2.5 pc (with $\mathcal{R}=1/4$); (c) 5 pc (with $\mathcal{R}=1/2$); (d) 10 pc (with $\mathcal{R}=1$). For all panels, the Mach number is 14. The fitting range is 1 pc to 10 pc, corresponding to the flat part of the energy spectrum of the hydrodynamical simulation (see figure \ref{powersim}).}\label{cvd_all}
\end{figure*}

Figure~\ref{cvd_all} shows the CVD-$L_{2D}$ relation for cases of various cloud thicknesses. The CVD-$L_{2D}$ relations with relative thickness $\mathcal{R}>1/10-1/8$ are nearly flat, consistent with the influence of thickness on the CVD-$L_{2D}$ relation discussed in last section.

It is likely that the centroid velocity of cores is affected little by the global cloud Mach number \citep[see also][]{Offner2009}. In fact, there is a no simple correspondence between the density distribution and the distribution of cores \citep{Padoan2002}. In some cases, the temperature, mean density, type of driving, scale of driving are as important or even more important. If the CVD is primarily inherited from the cloud turbulent velocities, then we would not expect the CVD power index to depend on the global velocity dispersion, which simply sets the normalization of the cloud energy.

\section{Conclusion and discussion}
\label{sec:conclusion}
We studied the relationship between the core velocity dispersion (CVD) and the projected separation of cores ($L_{2D}$)  in molecular clouds. We compared the results with an analytic model and hydrodynamic  simulations. We come to the following conclusions.

1) CVD is based on core centroid velocities, which reflects collective properties of dense gas condensations, thus is less affected by density fluctuations
\cite[c.f.][]{Brunt2004} and more sensitive to viewing geometry.

2) The relation between core velocities as a function of $L_{2D}$ and $l_{3D}$ is the same when the relative thickness $\mathcal{R}$ of a cloud (the ratio of cloud length to cloud depth along the line of sight) is smaller than 1/10 to 1/8. This result can potentially be used to constrain the line-of-sight cloud dimension, which is hard to measure directly. A simple functional fit to the CVD index and/or $h$ in equation \ref{CVD_2D} can potentially be used to provide a limit to the thickness (line of sight dimension) of a cloud.

3) The CVD-$L_{2D}(l_{3D})$ in the Taurus molecular cloud can be fitted by a power law with an index of 0.5, resembling Larson's linewidth-size relation. This may indicate that the $^{13}$CO cores in Taurus are distributed in a volume that is thinner than 2.5 pc-3.0 pc, corresponding to 1/10 to 1/8 of the transverse cloud scale. Performing the same analysis we conclude that the relative thickness $\mathcal{R}$ of Perseus and Ophiuchus is larger than 1/10-1/8.
The corresponding lower limits of the thicknesses are 1.6 pc-2.0 pc for Perseus, and 0.8 pc-1.0 pc for Ophiuchus.

4) We conclude that the thickness of a molecular cloud can be measured by using average line profiles in concentric regions of different sizes, where the $\sigma$-$L_{2D}$ relation flattens at small scales. The characteristic thickness $h$ of Taurus is 0.7 pc. The lower limits of the characteristic thicknesses $h$ of Perseus and Ophiuchus are 3.6 pc and 2.0 pc, respectively.

Our conclusions are consistent with priors observations, in which the scale and morphology of shells within the clouds indirectly constrains the cloud thickness \citep{Arce2011,Li2015}. This method provides a promising means of probing the cloud depth in regions where shells arising from stellar feedback are sparse or absent. In future works, we will apply this analysis to more diverse regions, e.g. molecular clouds with different column densities.

\acknowledgments This work is partly supported by  China Ministry of
Science and Technology under State Key Development Program for Basic
Research (2012CB821800, 2015CB857100) and the National Natural Science Foundation of China No. 11373038, No. 11373045, Strategic Priority Research Program of the Chinese Academy of Sciences, No. XDB09010302. SSRO acknowledges support from NASA through Hubble Fellowship grant \# 51311.01 awarded by the Space Telescope Science Institute, which is operated by the Association of Universities for Research in Astronomy, INC., for
NASA, under contract NAS 5-26555. The {\sc orion} calculations were performed on the Trestles XSEDE cluster and on the Yale University Omega cluster.


\end{document}